\documentclass[aps,prl,twocolumn,superscriptaddress,showpacs]{revtex4}
\usepackage{graphicx,bm}

\begin{document}
\title{Momentum Dependence of Charge Excitations in 
the Electron-Doped Superconductor Nd$_{1.85}$Ce$_{0.15}$CuO$_4$: a RIXS Study}
\author{K. Ishii}
\email{kenji@spring8.or.jp}
\affiliation{Synchrotron Radiation Research Center, Japan Atomic
Energy Research Institute, Hyogo 679-5148, Japan}
\author{K. Tsutsui}
\affiliation{Institute for Materials Research, Tohoku University,
Sendai 980-8577, Japan}
\author{Y. Endoh}
\affiliation{International Institute for Advanced Studies,
Kizugawadai, Kizu, Kyoto 619-0025, Japan}
\author{T. Tohyama}
\affiliation{Institute for Materials Research, Tohoku University,
Sendai 980-8577, Japan}
\author{S. Maekawa}
\affiliation{Institute for Materials Research, Tohoku University,
Sendai 980-8577, Japan}
\author{M. Hoesch}
\affiliation{Synchrotron Radiation Research Center, Japan Atomic
Energy Research Institute, Hyogo 679-5148, Japan}
\author{K. Kuzushita}
\affiliation{Synchrotron Radiation Research Center, Japan Atomic
Energy Research Institute, Hyogo 679-5148, Japan}
\author{M. Tsubota}
\affiliation{Synchrotron Radiation Research Center, Japan Atomic
Energy Research Institute, Hyogo 679-5148, Japan}
\author{T. Inami}
\affiliation{Synchrotron Radiation Research Center, Japan Atomic
Energy Research Institute, Hyogo 679-5148, Japan}
\author{J. Mizuki}
\affiliation{Synchrotron Radiation Research Center, Japan Atomic
Energy Research Institute, Hyogo 679-5148, Japan}
\author{Y. Murakami}
\affiliation{Synchrotron Radiation Research Center, Japan Atomic
Energy Research Institute, Hyogo 679-5148, Japan}
\affiliation{Department of Physics, Tohoku University, Sendai
980-8578, Japan}
\author{K. Yamada}
\affiliation{Institute for Materials Research, Tohoku University,
Sendai 980-8577, Japan}
\date{\today}

\begin{abstract}
We report a resonant inelastic x-ray scattering (RIXS) study of charge
excitations in the electron-doped high-$T_{\mathrm c}$ superconductor
Nd$_{1.85}$Ce$_{0.15}$CuO$_4$.  The intraband and interband excitations
across the Fermi energy are separated for the first time by tuning the
experimental conditions properly to measure charge excitations at low
energy.  A dispersion relation with ${\mathbf q}$-dependent width
emerges clearly in the intraband excitation, while the intensity of the
interband excitation is concentrated around 2~eV near the zone center.
The experimental results are consistent with theoretical calculation of
the RIXS spectra based on the Hubbard model.
\end{abstract}

\pacs{78.70.Ck, 74.25.Jb, 74.72.Jt}

\maketitle

The asymmetric features of the electronic phase diagram of the doping
dependence of the Mott insulating Cu oxides between hole- and
electron-doping have been an issue of debate for a long time. Their
exploration is very important for the understanding not only of the
mechanism of high-$T_{\mathrm c}$ superconductivity but also of the
effects of doping on a Mott insulator.  Experimental studies
investigating the reconstruction of the electronic bands by the carrier
doping have extensively been pursued~\cite{Uchida1,Onose2}, and now the
comprehension was reached that the Mott gap feature remains up to
considerable doping levels.  Recently angle-resolved photoemission
spectroscopy (ARPES) has provided plenty of information on the momentum
dependence of the occupied states~\cite{Damascelli1}. On the other hand,
the electronic band structure above the Fermi energy is still unclear
mainly due to the lack of convincing experimental data.  Electron energy
loss spectroscopy (EELS) suffers from multiple scattering at large
momentum transfers, while conventional optical methods, such as
photo-absorption, electronic Raman scattering and so forth, can observe
only the excitation at zero momentum transfer.  In this respect,
resonant inelastic x-ray scattering stands out as a unique and ideal
probe to be able to measure the momentum dependence of the electronic
excitations, in which the band structure of the unoccupied state is
elucidated through the two-particle excitation spectra. Current
experimental developments in the measurement of such charge dynamics or
electronic excitation has been directed to the doped Mott insulators, in
particular to a number of cuprates~\cite{Kim4,Hasan3}.

The Cu $K$-edge RIXS experiments of the parent compounds of the
high-$T_{\mathrm c}$ cuprates, such as Ca$_2$CuO$_2$Cl$_2$~\cite{Hasan1}
and La$_2$CuO$_4$~\cite{Kim1} showed a clear energy gap between the
occupied lower Hubbard band (LHB), more precisely the Zhang-Rice singlet
band, to the unoccupied upper Hubbard band (UHB).  Recent RIXS
measurements for the hole doped cuprates showed that the Mott gap
feature is robust even at considerable doping levels, besides
appreciable low energy excitations corresponding to the hole
doping~\cite{Kim4,Hasan3}.  In this letter, we extended the effort to
investigate the electron doped cuprates using RIXS and observe how the
Mott gap feature changes with the electron doping. Combined with an
extensive theoretical analysis, we can observe not only the interband
excitation across the Mott gap but also the intraband excitation within
the UHB throughout the whole Brillouin zone.

The RIXS experiments were carried out on the IXS spectrometer installed
at the beam line 11XU of SPring-8~\cite{Inami1}.  A Si (111)
double-crystal monochromator and a Si (400) channel-cut secondary
monochromator were utilized.  Horizontally scattered x-rays were
analyzed in energy by a bent Ge (733) crystal.  The overall energy
resolution is about 400 meV estimated from the full width at half
maximum (FWHM) of the quasielastic scattering.  Single crystals of
Nd$_{2-x}$Ce$_x$CuO$_4$ of $x$ = 0.15 and 0.075 were prepared, which
show superconductivity below $T_{\mathrm c}$ = 25 K, and
antiferromagnetic order below $T_{\mathrm N} \sim$ 120 K, respectively.
The surface of the crystal is normal to the {\it c}-axis, which was kept
in the scattering plane so as to be scanned in the reciprocal lattice
space spanned by either the [100]-[001] or the [110]-[001] axes.  All
spectra were collected at room temperature.

\begin{figure}
\includegraphics[width=7.cm]{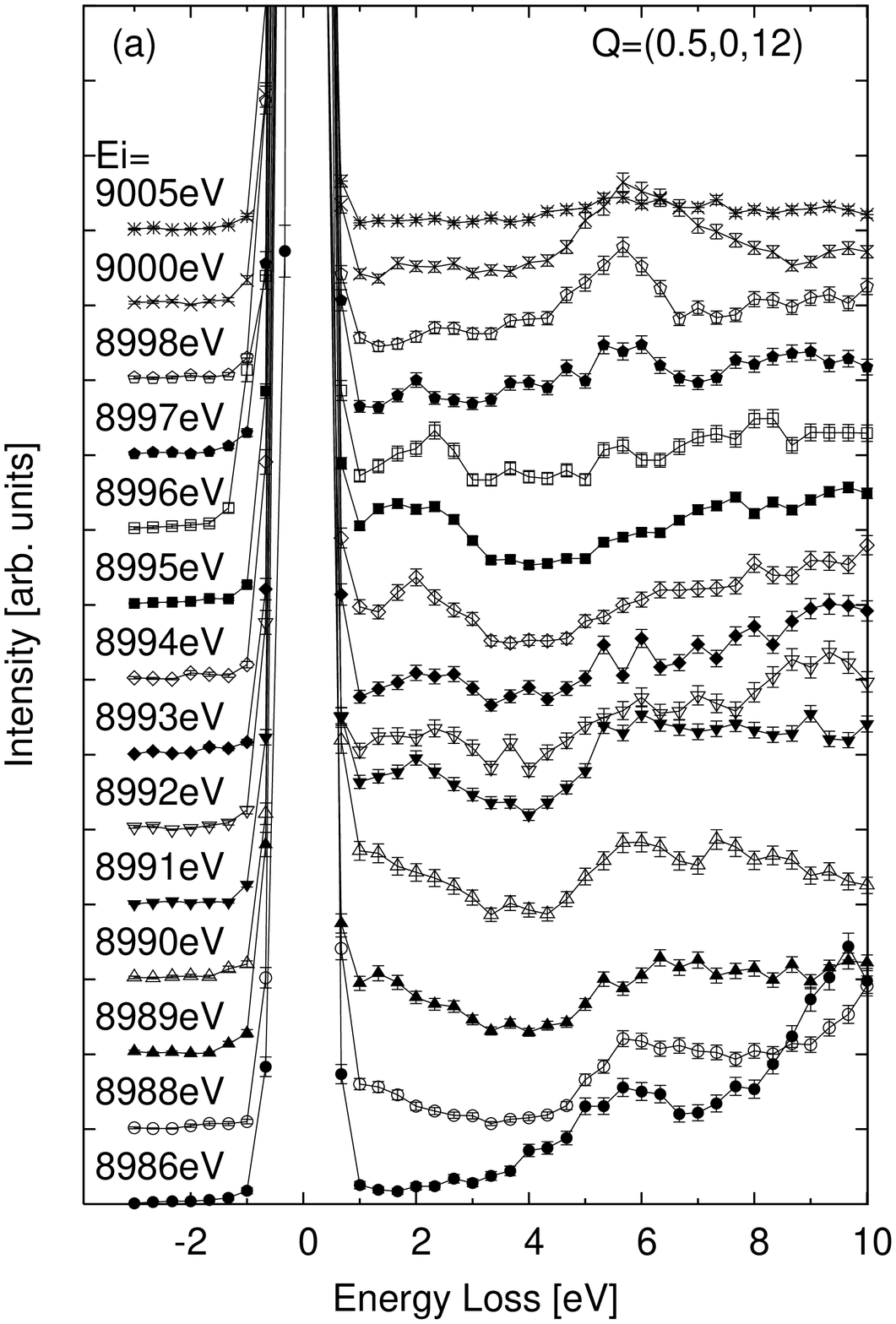}
\includegraphics[width=6.cm]{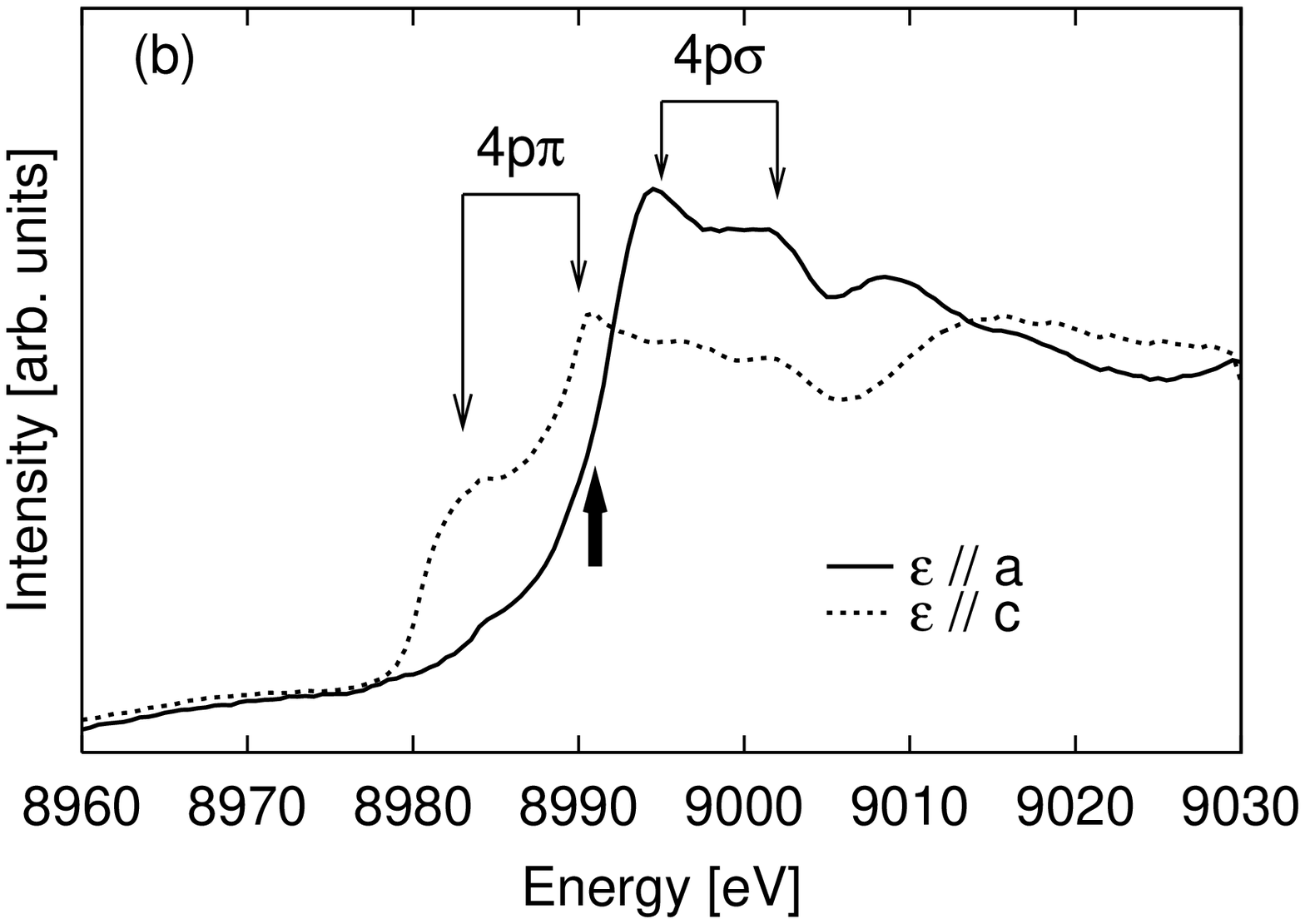}
\caption{(a) Resonant inelastic x-ray scattering spectra of
Nd$_{1.85}$Ce$_{0.15}$CuO$_4$ as a function of energy loss at some
representative incident x-ray energies $E_\mathrm{i}$.  The scattering
vector is fixed at $\mathbf{Q}=(0.5,0,12)$.  The strong intensity in the
spectra of $E_\mathrm{i}=8986$ eV at high excitation energy comes from
the Cu $K_{\beta 5}$ fluorescence. (b) X-ray absorption spectrum
measured by the fluorescence method.  The polarization of the x-rays
($\epsilon$) is parallel to the ${\mathbf a}$- or ${\mathbf c}$-axes of
the crystal.  The thick arrow indicates the energy used for
momentum-dependent RIXS measurements.}
\label{fig:eidep}
\end{figure}

Figures~\ref{fig:eidep} show the incident energy ($E_\mathrm{i}$)
dependence of RIXS, together with the fluorescence spectra.  The
absolute momentum transfer is fixed at $\mathbf{Q}=(0.5,0,12)$.
Resonantly enhanced peaks at around 2~eV and 6~eV can be seen.  The
latter peak was also observed in the undoped Nd$_2$CuO$_4$, and it was
identified as a charge transfer excitation to the antibonding
state~\cite{Hill1}.  We find two resonances at around 8990~eV and
9000~eV for the 6~eV peak.  These energies correspond to the resonance
energies for $\epsilon_\mathrm{i} \parallel c$ and $\epsilon_\mathrm{i}
\parallel ab$~\cite{Hamalainen1}, $\epsilon_\mathrm{i}$ being the
polarization of the incident x-ray.  Because the $\epsilon_\mathrm{i}$
is close to $\frac{{\hat \mathbf{a}}+{\hat \mathbf{c}}}{2}$ in our
experimental configuration of RIXS, where ${\hat \mathbf{a}}$ and ${\hat
\mathbf{c}}$ are the unit vectors along the $a$- and $c$-axes,
respectively, it is reasonable that both resonances are observed.  The
first two peaks in the fluorescence spectra are assigned to the
$1s-4p_{\pi}$ transition, while the next two are the $1s-4p_{\sigma}$
transition. The intensity of the 2~eV peak shows an enhancement at
around 8991~eV which corresponds to the absorption edge for
$4p_{\sigma}$.  Hereafter $E_\mathrm{i}$ is fixed at 8991~eV to focus on
the low energy excitations.

\begin{figure*}
\includegraphics[scale=0.3, trim=0 -5 0 0, clip]{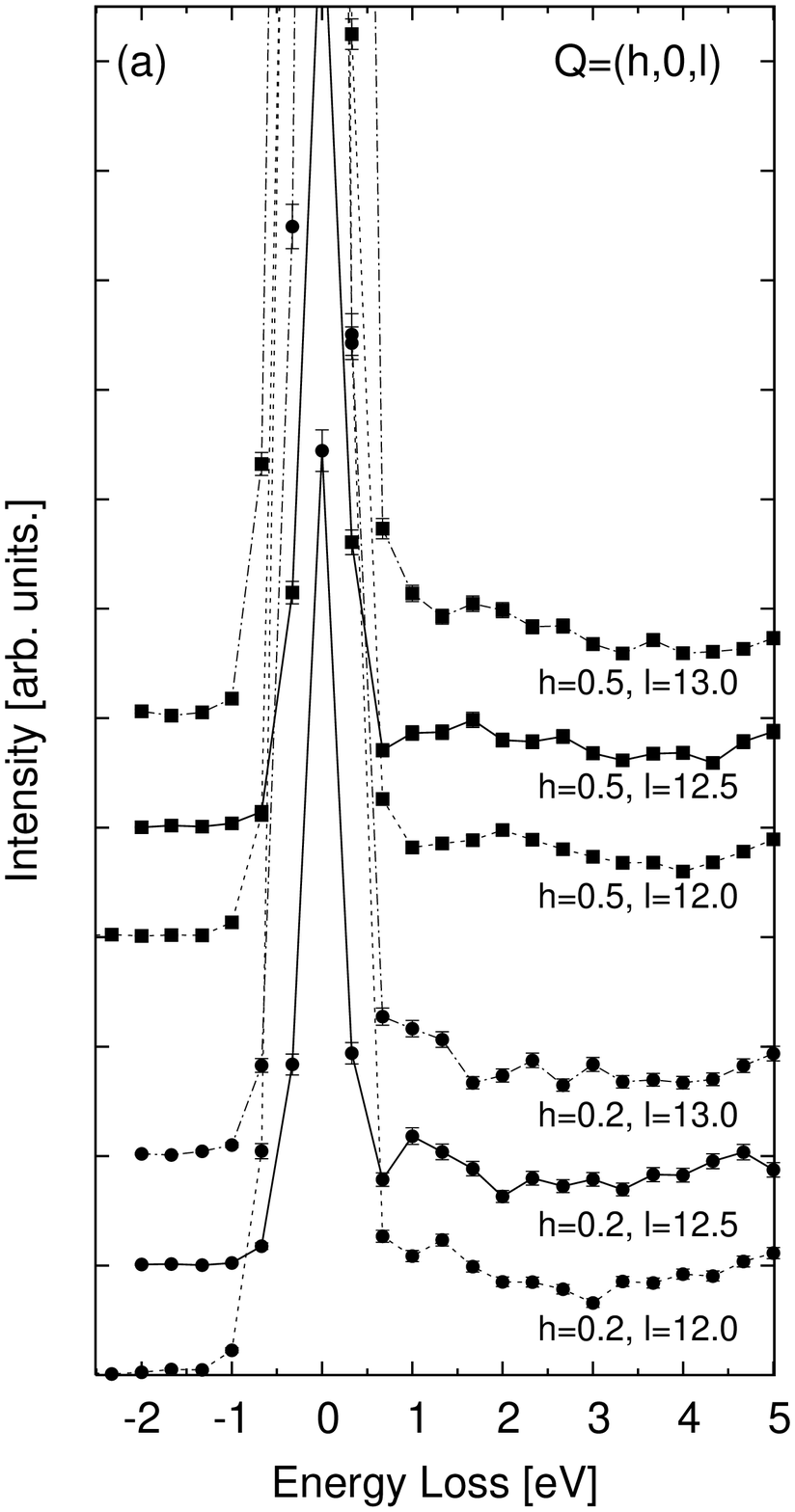}
\includegraphics[scale=0.3, trim=50 -5 0 0, clip]{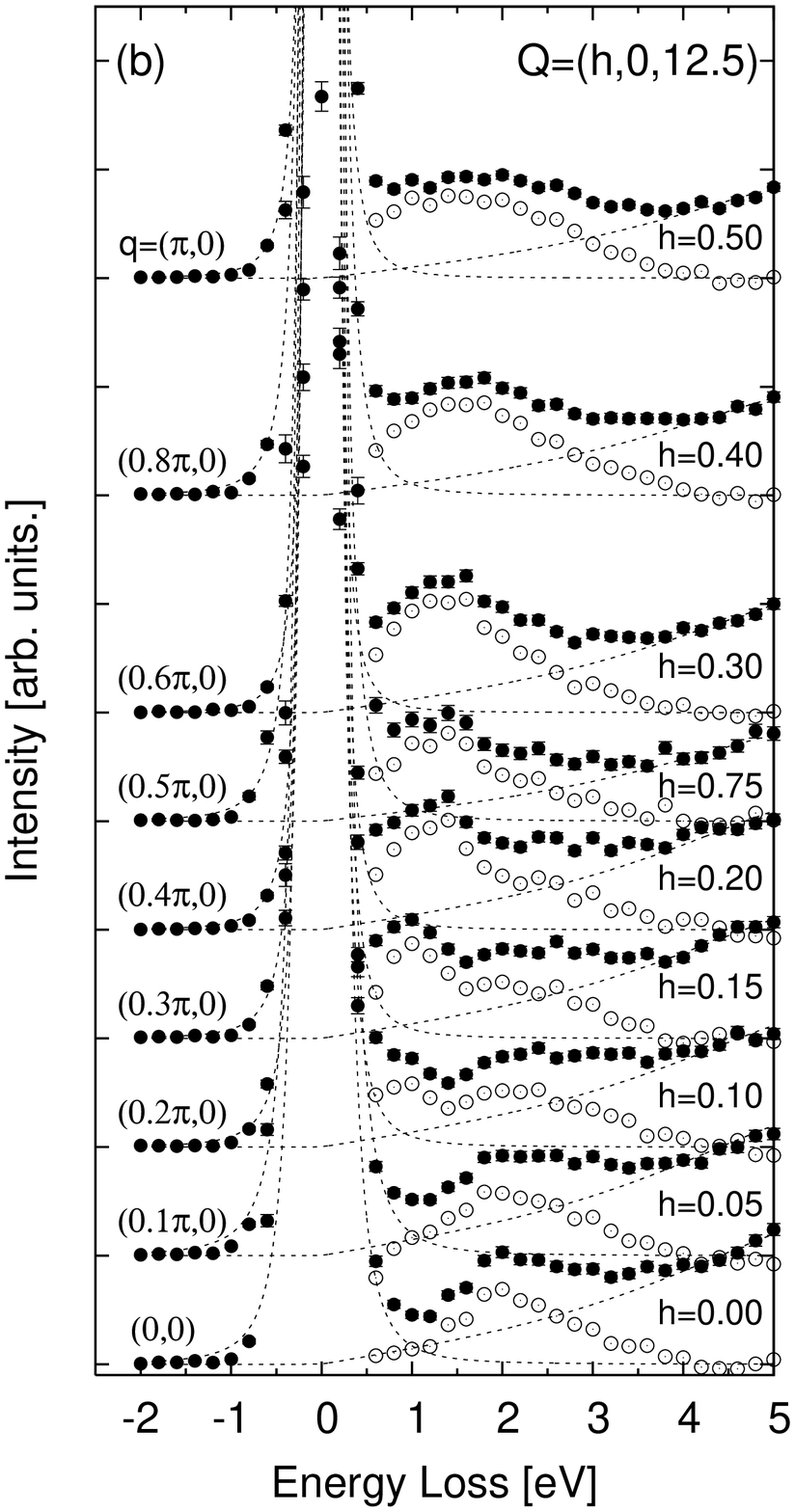}
\includegraphics[scale=0.3, trim=50 -5 0 0, clip]{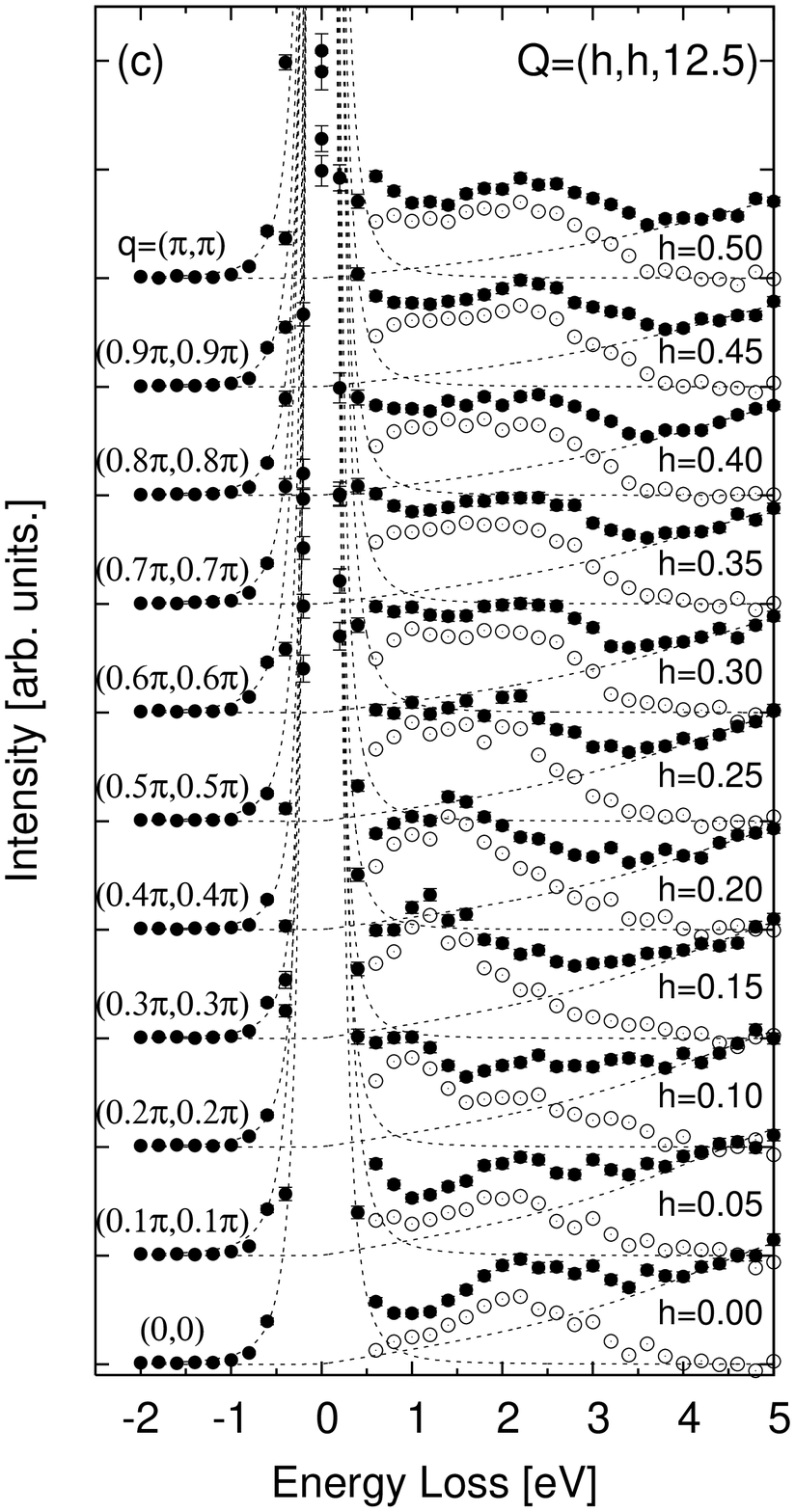}
\begin{minipage}[b]{0.25\linewidth}
\begin{center}
\includegraphics[width=5cm]{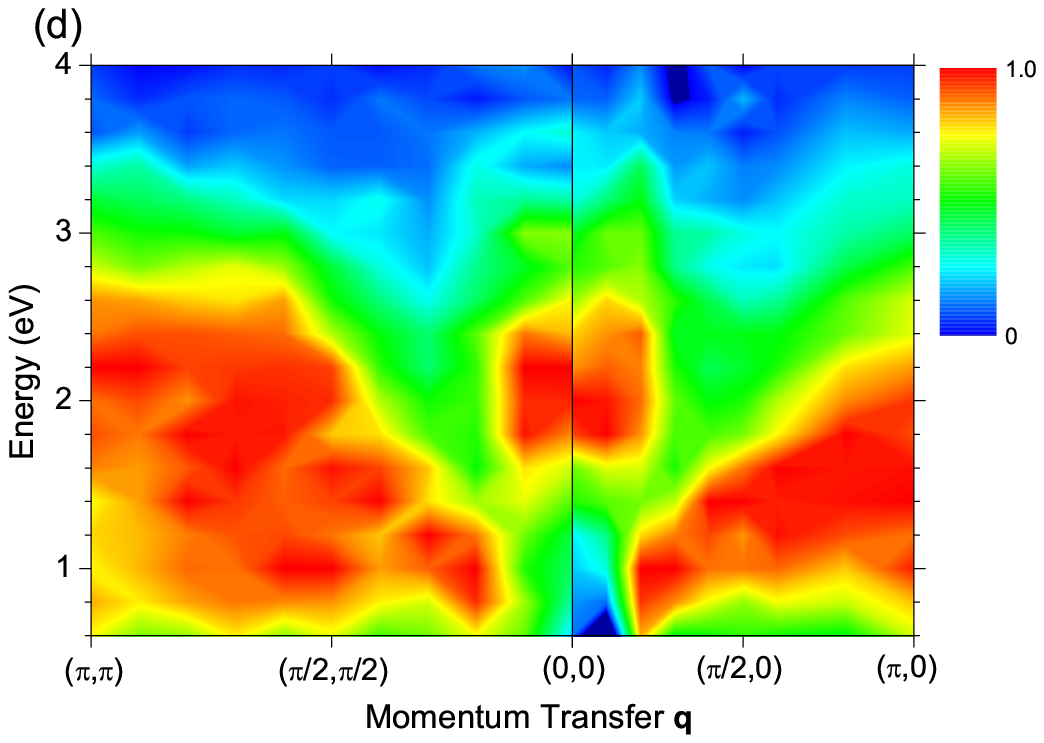}
\includegraphics[scale=0.3]{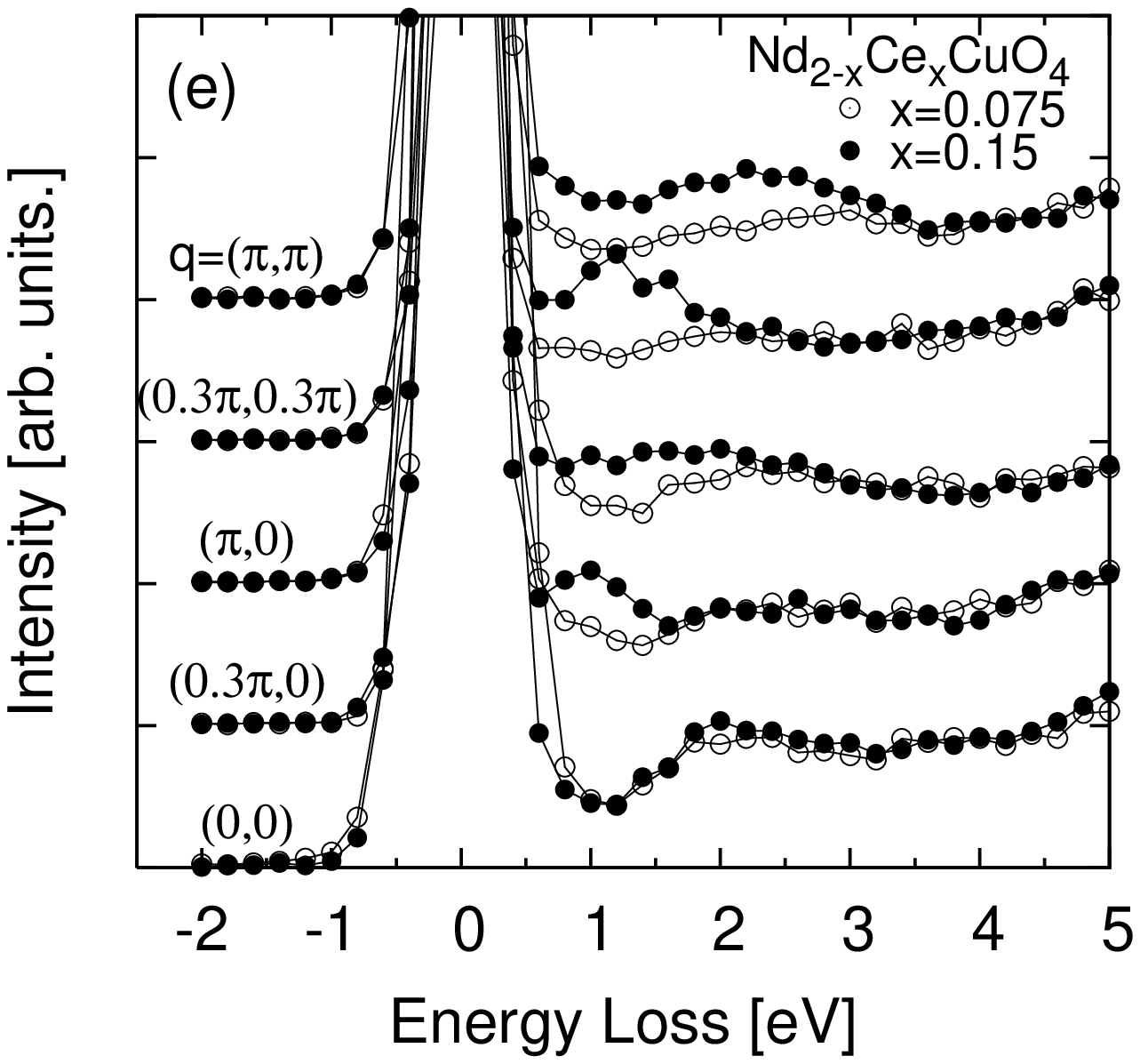}
\end{center}
\end{minipage}
\caption{Momentum dependence in Nd$_{1.85}$Ce$_{0.15}$CuO$_4$. (a) Along
the $c$-axis, (b) $(\pi,0)$, and (c) $(\pi,\pi)$ directions.
$E_\mathrm{i}=8991$~eV.  The filled symbols are raw data, and the open
ones in (b) and (c) are data from which the elastic scattering and the
scattering at higher energy (dotted lines) are subtracted. (d) Contour
plot of the RIXS intensity. After subtraction of the elastic and
high-energy contributions (open symbols in (b) and (c)) the data are
normalized for the maximum intensity in each momentum and interpolated
smoothly. (e) Comparison of the RIXS spectra to
Nd$_{1.925}$Ce$_{0.075}$CuO$_4$.}  \label{fig:qdep}
\end{figure*}

As shown in Fig.~\ref{fig:qdep}(a), the momentum dependence of RIXS
along the $c$-axis is weak, as expected from the strong
two-dimensionality of the CuO$_2$ plane.  However it should be noted
that the quasielastic tail is appreciably suppressed for the scan at
$l=12.5$.  In our experimental condition ($\pi$-polarization of the
incident x-ray), the intensity of the elastic scattering, whose major
component is Thomson scattering, is mostly proportional to $\cos^2
2\theta$, where $2\theta$ is the scattering angle and thus decreases
when $2\theta$ is close to 90~degree. It is crucially important to
reduce the elastic scattering to measure the low energy excitations in
RIXS.  For this reason, we selected $l=12.5$ to measure the momentum
dependence in the CuO$_2$ plane, though it is not a high symmetry plane.

Figures~\ref{fig:qdep} (b) and (c) show the momentum dependence of the
RIXS spectra along a line in the $\mathbf{q}=(\pi,0)$ and $(\pi,\pi)$
directions, respectively, where $\mathbf{q}$ represents the reduced
momentum transfer in the $ab$-plane.  Except for the spectrum at
$\mathbf{Q}=(0.75,0,12.5)$, all the spectra were measured at $h<0.5$ of
$\mathbf{Q}=(h,0,12.5)$ or $\mathbf{Q}=(h,h,12.5)$.  The spectrum at
$\mathbf{Q}=(0.75,0,12.5)$ lies between those of
$\mathbf{Q}=(0.2,0,12.5)$ and $\mathbf{Q}=(0.3,0,12.5)$, which indicates
that the electronic structure is symmetric with respect to $h=0.5$.  In
order to examine the momentum dependence more clearly, we subtract the
elastic contribution near 0~eV and the high-energy contribution above
4~eV from the raw data (dashed lines in Figs.~\ref{fig:qdep}(b) and
(c)).  The former is estimated from the anti-Stokes region, and the
latter is treated as a tail of the excitation at 6~eV by extrapolating
smoothly to the lower-energy region.  The open symbols in
Figs.~\ref{fig:qdep}(b) and (c) are the resulting spectra, where the
data below 0.6~eV are not shown due to the uncertainty in the assignment
of the quasielastic contributions. The spectra are replotted in
Fig.~\ref{fig:qdep}(d) as a contour map, where the maximum intensity at
each momentum point is normalized to unity and a smoothing procedure is
applied.  We can clearly see two characteristic excitations. One is the
excitation at 2~eV observed at the zone center. Its intensity rapidly
decreases with increasing $\mathbf{q}$. The other one is a broad but
dispersive excitation along the $(\pi,0)$ and $(\pi,\pi)$ directions.
As a function of $\vert \mathbf{q} \vert$, the latter excitation shifts
to higher energy up to 2-2.5~eV at the zone boundary, accompanied by an
increase of the spectral width. The upper edges of the excitations are
dispersive with a width of more than 2~eV from the zone center to the
zone boundary.

Two characteristic excitations just described above have been elucidated
further by duplicating the similar scans for $x=0.075$ crystal.  Typical
data of scans are shown in Fig.~\ref{fig:qdep}(e). The excitation
spectra at the zone center superpose each other and it is essentially
independent of $x$.  On the other hand, the spectra at finite
$\mathbf{q}$s show the weaker intensities for $x=0.075$ in lower-energy
region and the intensity seems to be proportional to $x$.  Such
dependence on doping indicates different nature between two excitations
which is identified by the following theoretical analysis of RIXS from
the electron-doped CuO$_2$ plane.

Keeping in mind the nearly monotonic dispersion like excitation mode, we
performed calculations of the RIXS spectrum using the numerically exact
diagonalization technique on a $4\times 4$ cluster of a Hubbard model
with the electron density $18/16=1.125$.  The model includes the hopping
of the electrons between first, second, and third nearest neighbor sites
($t$, $t'$, and $t''$, respectively) and the on-site Coulomb interaction
$U$.  The RIXS spectrum is expressed as a second-order process of the
dipole-transition between Cu $1s$ and $4p$ orbitals, where a Coulomb
interaction between a $1s$ core-hole and a $3d$ electron, $U_c$, is
explicitly included~\cite{Tsutsui1}.  We use $t'/t=-0.25$ and
$t''/t=0.12$, which are obtained from shape of the Fermi
surface~\cite{Armitage1}.  For other parameters, we take $U/t=8$,
$U_c/t=10$, and $t=0.3$~eV.  The inverse of the life time of the
intermediate state is assumed to be $\Gamma=3t$.

\begin{figure}
\includegraphics[width=8.5cm]{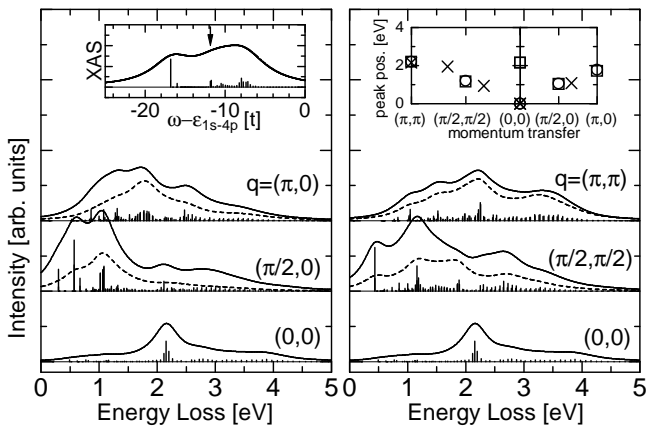}
\caption{The RIXS spectrum of an electron-doped $4\times 4$ Hubbard
cluster with long-range hopping terms.  The electron density is
$18/16=1.125$, and the model parameters are $t'/t=-0.25$, $t''/t=0.12$,
$U/t=8$, $U_c/t=10$, and $\Gamma/t=3$ with $t=0.3$~eV.  The
$\delta$-functions are convoluted with a Lorentzian broadening of
$0.8t$.  Broken lines: $N(\mathbf{q},\omega)$ for the same cluster.  The
inset of the left panel: the calculated absorption spectrum with a
Lorentzian broadening of $3t$. The vertical arrow denotes $E_\mathrm{i}$
for RIXS.  The inset of the right panel: momentum dependence of the peak
positions of RIXS (the squares), together with those of
$N(\mathbf{q},\omega)$ for $4\times 4$ (the circles) and
$\sqrt{18}\times\sqrt{18}$ (the crosses) clusters.}
\label{fig:theory}
\end{figure}

Figure~\ref{fig:theory} shows the calculated RIXS spectrum, where
$E_\mathrm{i}$ is set to a value denoted by the arrow in the absorption
spectrum shown in the inset.  The RIXS spectrum shows two characteristic
excitations similar to the observed ones: One is a 2~eV excitation at
$\mathbf{q}=(0,0)$, and the other is a broad band of excitations up to
$\sim3$~eV for all momenta except $(0,0)$.  The 2eV excitation is the
Mott gap excitation from LHB to UHB, as discussed in
Ref.~\cite{Tsutsui2}.  The broad excitations can be assigned to the
intraband excitation.  To confirm this, we calculated the dynamical
density response function $N(\mathbf{q},\omega)$, which can describe the
momentum-dependent intraband and interband density fluctuations
separately when $U$ is
large~\cite{Tohyama1,Eder1,Khaliullin1,DHKim1,Grober1}.  Comparing RIXS
and $N(\mathbf{q},\omega)$ (broken lines), we find qualitatively similar
behavior for all momentum and energy regions except for the 2~eV
excitation at $\mathbf{q}=(0,0)$.  This means that the broad and
dispersive excitations observed in Fig.~\ref{fig:qdep} come from the
charge fluctuations in the metallic phase.

The energy position of the highest peak in RIXS and
$N(\mathbf{q},\omega)$ is plotted in the inset of the right panel of
Fig.~\ref{fig:theory}.  In addition to the positions taken from the main
panels, we plot peak positions obtained from a
$\sqrt{18}\times\sqrt{18}$ cluster with an electron density $20/18$.
The resulting momentum dependence is found to trace out the center the
of broad spectra in the contour plot of Fig.~\ref{fig:qdep}(d).  This
agreement justifies the assignment of the observed two structures to the
Mott-gap and intraband excitations.

As emphasized above, the intraband excitations show broad features. Such
a broadness comes from strong correlation common to Hubbard-type
models~\cite{Tohyama1,Eder1,Khaliullin1,DHKim1,Grober1}, and thus it is
independent of the presence of the long-range hoppings.  Instead, the
effect of $t'$ and $t''$ may appear in the very low-energy region, where
excitations are predominately controlled by the Fermi surface topology
inducing $2k_\mathrm{F}$ and $4k_\mathrm{F}$ branches~\cite{Tohyama1},
$k_\mathrm{F}$ being the Fermi momentum.  Unfortunately, at present the
energy resolution of RIXS is limited so that it is difficult to resolve
the branches.  We continue our efforts to improve the energy
resolution and this will open a new view of the intraband charge
excitations in the high-$T_c$ cuprates in the future.

Although we made use of electron-doped Nd$_{1.85}$Ce$_{0.15}$CuO$_4$, it
may also be possible to use hole-doped materials such as
La$_{2-x}$Sr$_{x}$CuO$_4$ to detect the intraband excitation. However,
it is important to notice that there is an advantage of electron doping
over hole doping.  As seen in the inset of the left panel of
Fig.~\ref{fig:theory}, the absorption spectrum shows three components:
$\omega-\epsilon_{\mathrm{1s-4p}}=-17t$, $-12t$, and $-8t$.  The latter
two components are also seen in the undoped system.  On the other hand,
the former appears upon electron doping only, corresponding to a final
state where the core hole attracts a doped electron on the same site.
Therefore, the final state hardly contains any pair of empty and doubly
occupied sites, i.e., no excitations across the Mott gap.  This means
that, the intraband charge excitations dominate the RIXS spectrum if
$E_\mathrm{i}$ is tuned to the lowest-energy peak.  As long as we fix
$E_\mathrm{i}$ to the absorption-edge region, the contribution from the
lowest-energy peak induces large intraband charge excitations.  In our
case we select the incident energy to the absorption edge for
$4p_{\sigma}$ as shown in Fig.~\ref{fig:eidep}(b), and this condition is
satisfied.  In contrast, a corresponding final state in the hole-doped
system, which emerges after hole doping, exists in a much higher-energy
region at around $\omega-\epsilon_{\mathrm{1s-4p}}=0$~\cite{Tsutsui2}.
Experimentally tuning the incident photon energy to this region seems to be
difficult because of the overlap of other absorption processes.

Finally, we compare the interband excitations across the Mott gap
between hole- and electron-doping. Recently RIXS experiments for
La$_{2-x}$Sr$_{x}$CuO$_4$ have been reported~\cite{Kim4,Hasan3}. Their
results showed that the spectral shape is almost independent of the
momentum transfer except for small shifts in energy. On the other hand,
the interband excitation of Nd$_{1.85}$Ce$_{0.15}$CuO$_4$ concentrates
on a energy ($\sim 2$~eV) at the zone center and becomes broad in energy
with increasing momentum transfer.  Such a difference in momentum
dependence is consistent with a previous theoretical
result~\cite{Tsutsui2}, where the difference in the strength of
antiferromagnetic correlation plays a crucial role.

In summary, we have performed a RIXS study for the electron-doped
superconductor Nd$_{1.85}$Ce$_{0.15}$CuO$_4$, and found
characteristics of the intraband and interband excitations. The
intraband excitation shifts to higher energy with the increase of the
peak width as a function of momentum transfer.  The spectral shape of
the intraband excitation has a similarity to $N(\mathbf{q},\omega)$ of
the two-dimensional Hubbard model. This demonstrates that RIXS is a
good tool to measure momentum-dependent density fluctuations in
strongly correlated metallic systems.  On the other hand, the
interband excitation across the Mott gap is enhanced in intensity at
the zone center, which is in contrast to the momentum-independent
spectral shape in hole-doped La$_{2-x}$Sr$_x$CuO$_4$.

The authors thank T. Uefuji for supplying a crystal of
Nd$_{1.925}$Ce$_{0.075}$CuO$_4$.  K. T., T. T., S. M., and K. Y. were
supported by the Japanese Ministry of Education, Culture, Sports, Science
and Technology, Grant-in-Aid for Scientific Research.  K. T., T. T., and
S. M. were also supported by CREST, NAREGI Nanoscience Project.
M. H. acknowledges support from the Japanese Society for the Promotion
of Science.  The numerical calculations were partially performed in the
supercomputing facilities of ISSP, University of Tokyo and IMR, Tohoku
University.

\bibliography{ixs,cuprate}
\end{document}